\documentclass[conference]{IEEEtran}

\usepackage{cite}
\usepackage{amsmath,amssymb,amsfonts}
\usepackage{algorithm}
\usepackage{algorithmic}
\usepackage{graphicx}
\usepackage{textcomp}
\usepackage{xcolor}
\usepackage{booktabs}
\usepackage{hyperref}
\usepackage{tabularx}

\usepackage{chngcntr}
\counterwithout{figure}{section}
\counterwithout{table}{section}

\hypersetup{
    colorlinks=true,
    linkcolor=black,
    citecolor=black,
    urlcolor=blue
}

\title{Space-based AI Infrastructure: Could it Fly?}
\title{The Cost and Network Limits \\ of Space‑Based AI Compute}
\author{
\IEEEauthorblockN{Kees van Berkel\IEEEauthorrefmark{1} }
\IEEEauthorblockA{\IEEEauthorrefmark{1}
	Technical University Eindhoven, Dept. of Math. \& Computer Science \\
	 P.O. Box 513, 5600 MB Eindhoven, The Netherlands \\
	Email: c.h.v.berkel@tue.nl}
}

\begin{document}

\maketitle


\begin{abstract}
This paper evaluates whether large‑scale AI data centers deployed in low‑Earth orbit (LEO) 
could become a cost‑effective alternative to terrestrial facilities. 
The analysis compares orbital and ground‑based systems across launch cost, power generation, cooling, 
radiation exposure, and atmospheric reentry, as well as compute‑network performance. 
A key distinction is the shift from terrestrial Clos networks to space‑based mesh networks using laser inter‑satellite links. 
Using bisection bandwidth, bisection intensity, and roofline‑style models, 
we show that while LEO‑based inference may be feasible, 
training frontier‑scale LLMs in orbit is unlikely to be competitive with terrestrial data centers.  
\end{abstract}



\section{Introduction}

\footnotetext[0]{The author is also a consultant to Snap Inc, Eindhoven.}
As terrestrial compute approaches hard limits in energy, cooling, and land use, 
orbital AI is emerging as a credible next frontier.
The vision behind orbital data centers was first articulated in a white paper by a startup named Starcloud 
\cite{2024-starcloud-spaceai}. 
As benefits they claim inexpensive solar energy, lower cooling costs using passive radiative cooling, 
and the opportunity to scale such orbital data centers almost indefinitely.

In October 2025 Elon Musk's shared his bold view on orbital AI data centers \cite{2025-xai-joins-spacex}: 
\begin{quote}
	``The basic math is that  launching  a million tons per year of satellites 
	generating  100kW of compute power per ton
	would add  100 gigawatts of AI compute capacity annually, 
	with no ongoing operational or maintenance needs.''  \\
	``My estimate is that within 2 to 3 years, 
	the lowest-cost way to generate AI compute will be in space.'' 
\end{quote}
Scalability in Musk's wording: ``Ultimately, there is a path to launching  1TW/year from Earth.'' 
In February 2026 SpaceX applied for  FCC permission to put up to a million data center satellites in orbit
\cite{2026-spacex-orbitaldatacenter}. 
The SEC filing by SpaceX \cite{2026-spacex-secfiling} in May 2026 is also a useful source of technical detail.


\subsection{Problem Statement}

This paper addresses two central questions\footnote{
	This  paper is based on the MPSoC'26 presentation ``AI in The Sky =? Pie in The Sky?",
	June 23, Victoria, Canada.}.
\begin{enumerate}
\item How would a 1‑GW orbital AI constellation perform on training and inference workloads 
	compared to a 1‑GW terrestrial data center?
\item How will technology trends -- satellite mass, compute density, and network bandwidth --
	affect a possible performance gap over the next five years?
\end{enumerate}

The main perspectives on these two questions are those of capital expenditures (CapEx) 
and operational expenses (OpEx). 
Here we do not aim at absolute numbers, such as tokens/\$, 
but at relative factors between costs in space and costs on earth.
Hence, the analysis focuses on the {\em differences} between 
the construction and operation of a terrestrial AI data center versus those of an orbital AI data center.
These differences fall into two broad categories:
\begin{enumerate}
 \item {\bf physics‑related constraints} (Section \ref{sec:physics}), 
 	such as launch cost, power, cooling, radiation tolerance, electromagnetic emission, and atmospheric reentry; 
 \item {\bf compute‑related constraints} (Section \ref{sec:compute}), 
 	especially inter‑satellite communication.
\end{enumerate}


\subsection{Related Work}

Tsinghua University proposes a hierarchical space-based computing network architecture 
\cite {2025-tsinghua-spacebased}.
Its description is mostly qualitative and the few numbers mentioned fall orders of magnitude short
of those needed for AI data centers.

Hangzhou University offers an overview of space-computing architectures, 
including a description of the key technologies needed \cite{2025-hangzhou-spacebased}. 
They describe the need for {\em distributed collaborative training and inference}
under limited resource conditions.
While this idea sounds intriguing they also state that 
{\em research in space collaborative computing is still in a nascent stage}.
The paper lacks a quantitative analysis.

Google proposes a planar 81-satellite constellation placed at a mean cluster altitude of 650 km,
with a cluster radius of 1 km \cite{2025-aguera-spaceai}.
They discuss the orbital dynamics of such a cluster, 
results on TPU (Google's AI chip) radiation testing, as well as high-bandwidth inter-satellite links, 
The latter two will be revisited in Section \ref{sec:discussion}. 
The paper does not offer detail on the network topology of the satellite constellation or its performance.


\section{Background}
\label{sec:background}


\subsection{Terrestrial AI Data Centers}
\label{subsec:data-centers}

A modern AI data center is organized as a layered compute and network hierarchy.
At the base of the hierarchy is this GPU, 
the fundamental compute element that executes the tensor‑core operations driving large‑scale LLM training and inference.

A {\em server} integrates several tightly-coupled GPUs, typically 4 to 8, 
connected by an intra‑server GPU fabric such as NVLink or NVSwitch. 
A {\em rack} aggregates dozens of these servers 
and provides a switch domain with predictable bandwidth and low‑diameter connectivity.
A state-of-the-art rack delivers about half an ExaFLOP/s ($10^{18}$ 8b-16b floating-point operations),
while consuming 120kW \cite{2024-nvidia-blackwell}.

A {\em pod} groups multiple racks into a high‑bisection cluster, usually built with a Clos topology \cite{2015-google-clos}, 
and forms the largest unit that can sustain near‑uniform all‑to‑all communication for training.
The {\em multi‑pod fabric} connects many pods across a datacenter, 
trading raw bandwidth for scale and resilience. 
It supports global job scheduling, checkpointing, and distributed training patterns.
A 1‑GW facility constructed from 8,000 racks (120kW each) delivers  about $10^{22}$ FLOP/s.

A terrestrial 1\,GW AI data center may cost 35-60B\$ to construct,
and about 1B\$ a year to operate. See Table \ref{tab:data-centers},
based on data from Epoch AI \cite{2026-EpochAI-DataCenterCost} and NVIDIA.

\begin{table}[h!]
	\centering
	\begin{tabular}{lrr}												\hline
		 \textbf{CapEx}		& \textbf{Cost [B\$]} 		& \textbf{fraction [\%]} \\	\hline
		Compute hardware 		& 12--20 			& 35--40 	\\
		Electrical infrastructure 	& 14--18 			& 40--45 	\\
		Cooling systems 		& 5--10 			& 15--20 	\\
		Land + construction 		& 7--12 			& 10--15 	\\
		Networking 			& 1--2 			& 3--5 	\\
		Water + utilities 		& 0.5--1 			& 1--3 	\\
		Total					& 35--60			& 100  	\\			\hline
	\end{tabular} \\
	\vspace{2mm}
	\begin{tabular}{lrr}												\hline
		 \textbf{OpEx}	& \textbf{Annual Cost [B\$]} 	& \textbf{fraction [\%]} \\	\hline
		Energy 				& 0.6 			& 65--70 	\\
		Labor 				& 0.1--0.2 			& 10--20 	\\
		Maintenance 			& 0.1 			& 10--12 	\\
		Taxes + land 			& 0.05--0.1 		& 5--8	\\
		Total					& 0.9				& 100 	\\			\hline 
	\end{tabular}
	\vspace{1mm}
	\caption{Capex and opex for high-end terrestrial AI data centers \cite{2026-EpochAI-DataCenterCost}.}
\label{tab:data-centers}
\end{table}

The {\em data center as a computer} \cite{2025-datacenter-as-computer}  perspective
treats the entire AI datacenter as a single, unified machine rather than a collection of independent servers. 
It reframes racks, pods, networks, and accelerators as architectural components, 
much like cores, caches, and buses inside a traditional computer.
		

\subsection{AI Training and Inference}
\label{subsec:models}

Large language models are built on deep neural networks 
composed of many interconnected {\em  layers} that transform incoming data. 
Modern LLMs predominantly employ the transformer architecture, 
whose attention mechanisms enable the model to capture long‑range dependencies within text. 

During training, the model optimizes a vast number of {\em parameters} — often in the billions or trillions — 
that encode statistical regularities of language. 
Text is represented as discrete {\em tokens} arranged into ordered token sequences, 
which the model processes to learn the structure of (natural) language.

At inference time, the trained model applies these learned parameters to new token sequences, 
using its internal representations to predict subsequent tokens and generate coherent, contextually appropriate output. 

The training of a large language model 
requires the distribution of the computation across hundreds of thousands of GPU accelerators. 
This distribution relies on several complementary forms of parallel computing \cite{2025-UCal-WLB-LLM}: 
\begin{itemize}
\item {\em Data parallelism} replicates the full model across multiple servers or pods, allowing each replica to process different minibatches while synchronizing gradients to maintain a consistent parameter state. 
\item {\em Tensor parallelism} partitions individual matrix operations across accelerators, enabling layers whose parameter or activation sizes exceed the memory or compute capacity of a single device. 
\item {\em Pipeline parallelism} divides the model’s layers into sequential stages distributed across servers or racks, 
with micro-batches flowing through the pipeline to keep all stages active. 
\item Additional strategies such as {\em sequence parallelism} and {\em expert parallelism}
further decompose computation along the token dimension or across sparsely activated expert networks. 
\end{itemize}

A high token throughput requires a balanced combination of these forms of parallelism.
This balance depends on the LLM model size, the machine size, the communication network, 
and  the forms and volumes of the input data \cite{2025-UCal-WLB-LLM}. 
The amount of parallelism achieved during the training of frontier LLM models
is measured as {\em model flop utilization} (MFU) and has reached truly impressive levels of 40\% to 70\%
\cite{2021-mfu-nvidia, 
	2022-mfu-training-google, 
	2022-mfu-nvidia, 
	2023-mfu-nvidia, 
	2024-mfu-bytedance,
	2024-mfu-meta}.
During each clock cycle (at roughly 2GHz), as many as 40\% - 70\% of the GPU multiply-accumulate (MAC) units contribute to the LLM training.

\begin{figure}[t]
\centering
    \includegraphics[width=\linewidth]{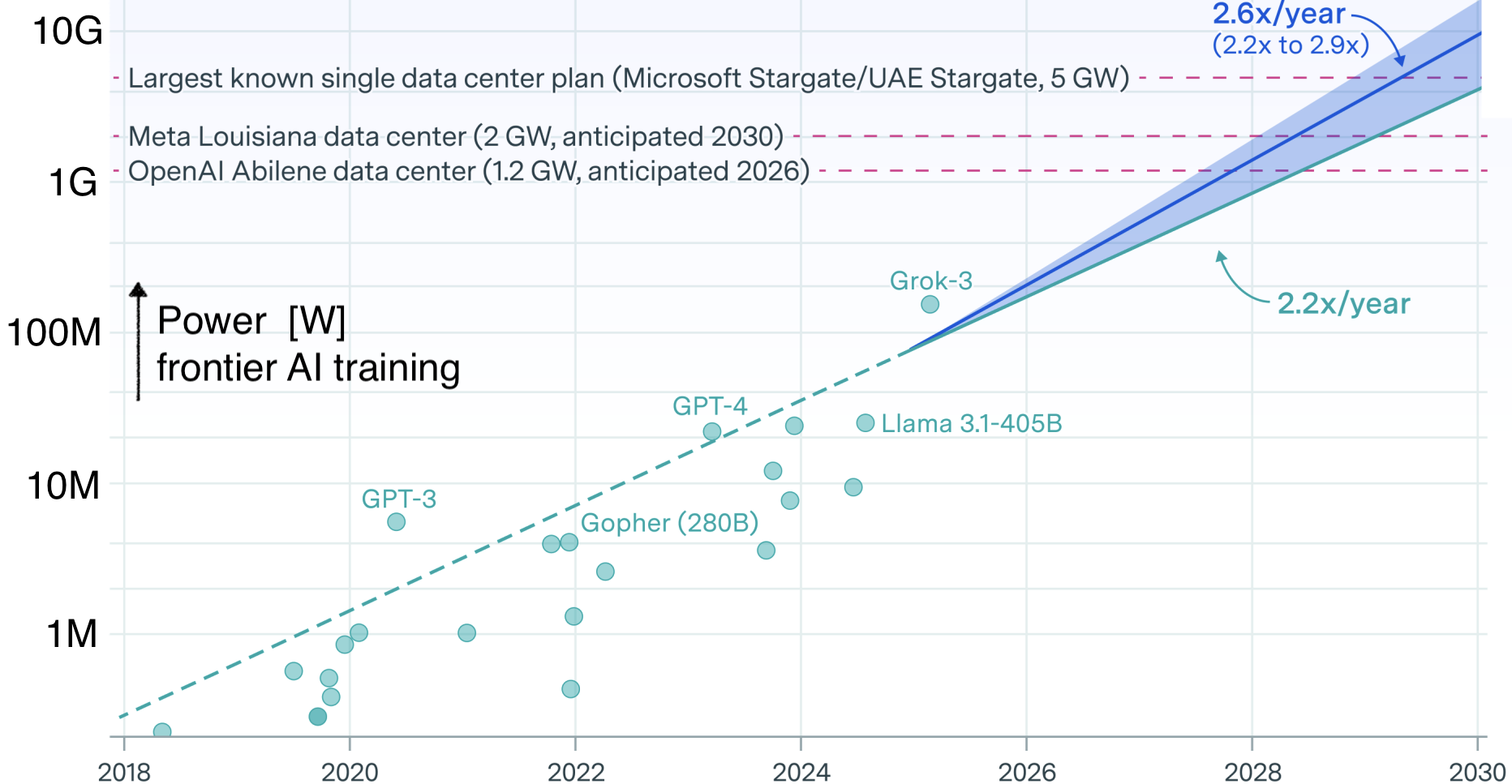}
    \caption{Projected power growth for frontier AI training, from \cite{2025-epri-scaling-intelligence}.}
\label{fig:training-power}
\end{figure}

Training frontier‑scale LLMs typically requires several months of continuous operation.
The power consumption during that training is measured in tens of MW, and is increasing at a steady rate of
$2.2\times$ per year, as shown Figure \ref{fig:training-power}.

\begin{figure}[t]
   \centering
    \includegraphics[width=0.8\linewidth]{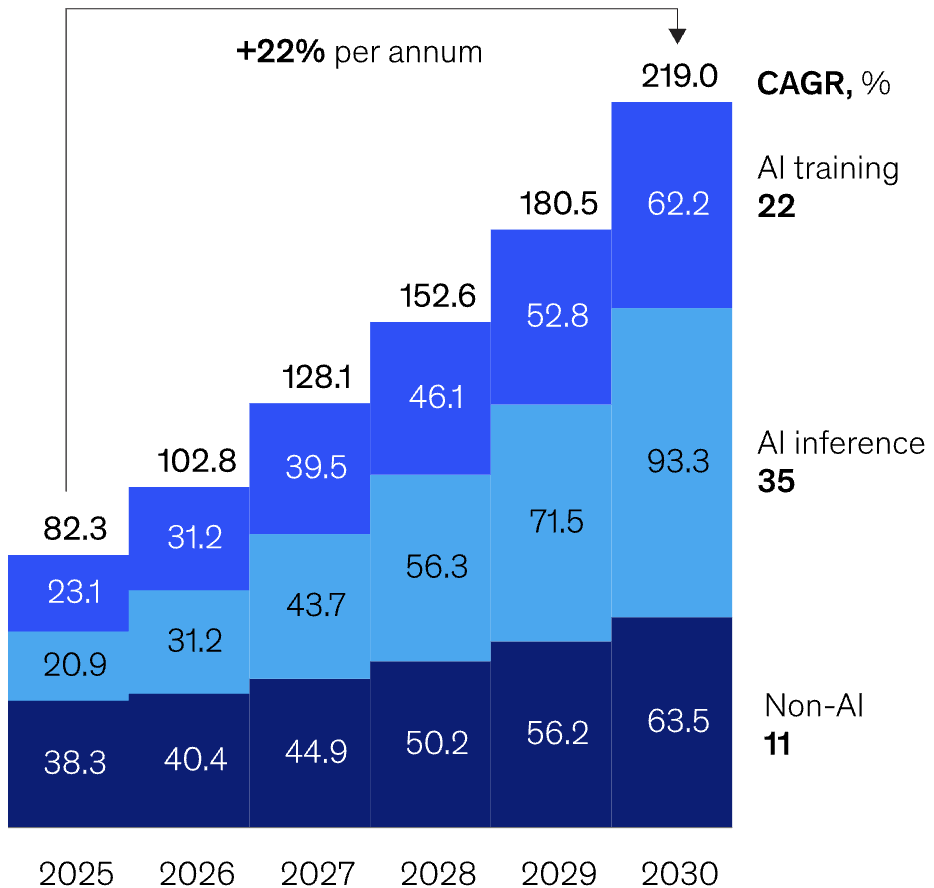}
   \caption{The annual global investment in data centers, in GW (from \cite{2025-mckinsey-next-big-shifts}).}
\label{fig:data-center-investment}
\end{figure}

The projected annual global investment in data centers is shown in Figure \ref{fig:data-center-investment}.
Note that the compound annual growth rates (CAGR) for AI training (1.22$\times$) 
is substantially less than the 2.1$\times$ growth in training power consumption.
This apparently implies that in the future LLM training is applied to ever larger, but fewer models per year.
Also note that in 2030 the investments both in AI training and in AI inference are well below Musk's 100GW per year,
as highlighted in the introduction.


\subsection{Performance Models for AI Compute}
\label{performance-models}

Performance models are used to analyze and predict the compute performance of AI data centers,
and to optimize the balance of the aforementioned forms of parallelism.
These performance models can be divided in three broad classes: analytical, profiling-based, and simulation-based. 
See \cite{2026-imec-performance-model} for an overview of these classes, including references. 



IMEC describes a performance model \cite{2024-imec-performance-model} that includes the
costs of network communication due to {\em all-reduce} operations, common in LLM training and inference.
The overall computation time is expressed as
\begin{equation}
	\begin{array}{rllllll}
	    T 	& = 	& T_{\textit{latency}} 	& +	& T_{\textit{bandwidth}}		& +	&  T_{\textit{compute}}  	\\[1mm]
		& = 	& 2 L \log(N)		& +	& \dfrac{2K}{N \times B}(N – 1) 	& + 	& \gamma\dfrac1N ~,	\\
	\end{array}		
	\label{eq:imec-model}
\end{equation}
where  $N$ is the number of compute nodes, $K$ the data volume, $L$ the network latency, $B$ the network bandwidth,
and $\gamma$ a factor to weigh the relative contribution of compute.
The $\log$ factor in the latency term stems from the tree-like topology of the communication network.
The network bandwidth $B$ is loosely defined as ``the network bandwidth between processors''.
They observe that for training the contribution of the latency term is negligible, unlike for inference.

\begin{figure}[t]
   \centering
    \includegraphics[width=0.9\linewidth]{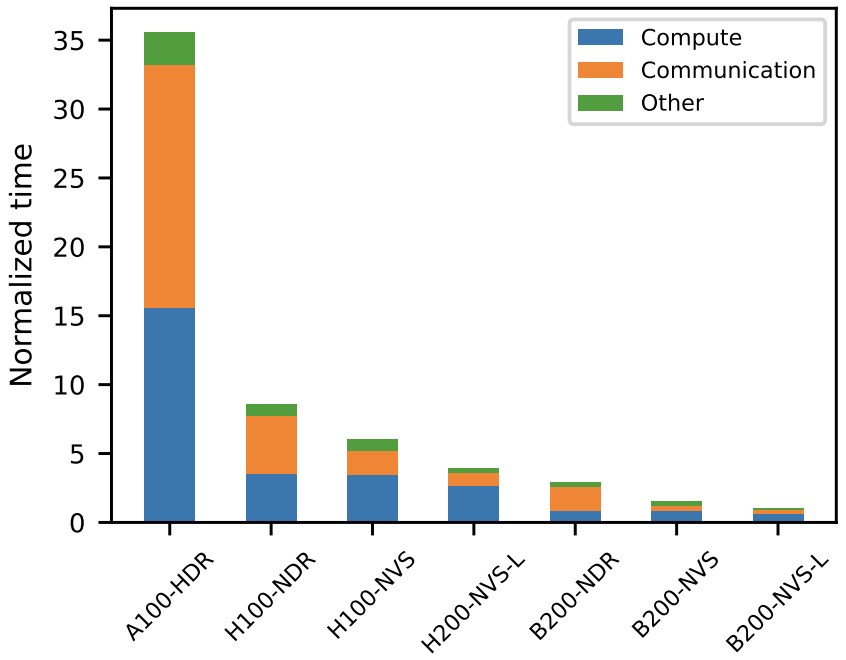}
    \caption{A breakdown of normalized training times across multiple GPU generations for GPT-3, 175B parameters
    (from \cite{2024-imec-performance-model}).}
\label{fig:imec-model}
\end{figure}

Figure \ref{fig:imec-model} shows how the breakdown of the total computation times varies across multiple GPU generations for GPT-3 (175B parameters). Significantly, the share of the communication time (dominated by $T_{\textit{bandwidth}}$) varies between 20\% and 50\%.
Apparently, there are good reasons why AI data centers have such very dense, high bandwidth, communication networks.

		
\subsection{LEO Satellite Networks}
\label{subsec:starlink}

Starlink is a satellite constellation that provides broadband internet service across the world \cite{2026-wikipedia-starlink}.
To date, over 10,000 satellites are in Low Earth Orbit (LEO, $\approx 500-600$km),
with another 12,000 satellites planned.
Each launch of a Falcon 9 rocket brings an additional 20-29 Starlink V2 Mini satellites into LEO.
During 2025, SpaceX launched 12-16 Falcons per month.

The Starlink satellites together form a  communication network, 
based on laser inter-satellite links \cite{2021-CarletonU-starlink-lisl} with a bandwidth of up to 100Gb/s.
Each satellite can communicate with a handful of other satellites in parallel, using steerable laser beams.
Starlink forms the largest communication infrastructure in space, and is used as a baseline for this study.
More detail on Starlink is introduced in Section \ref{sec:physics}.


\section{The Physics Side of Orbital AI}
\label{sec:physics}

This section addresses the costs of orbital AI from a physics perspective.
Section \ref{sec:compute} discusses the compute  perspective.


\subsection{Cost of Launch to Low Earth Orbit}
\label{subsec:cost-to-leo}

\begin{figure}[t]
    \centering
    \includegraphics[width=0.9\linewidth]{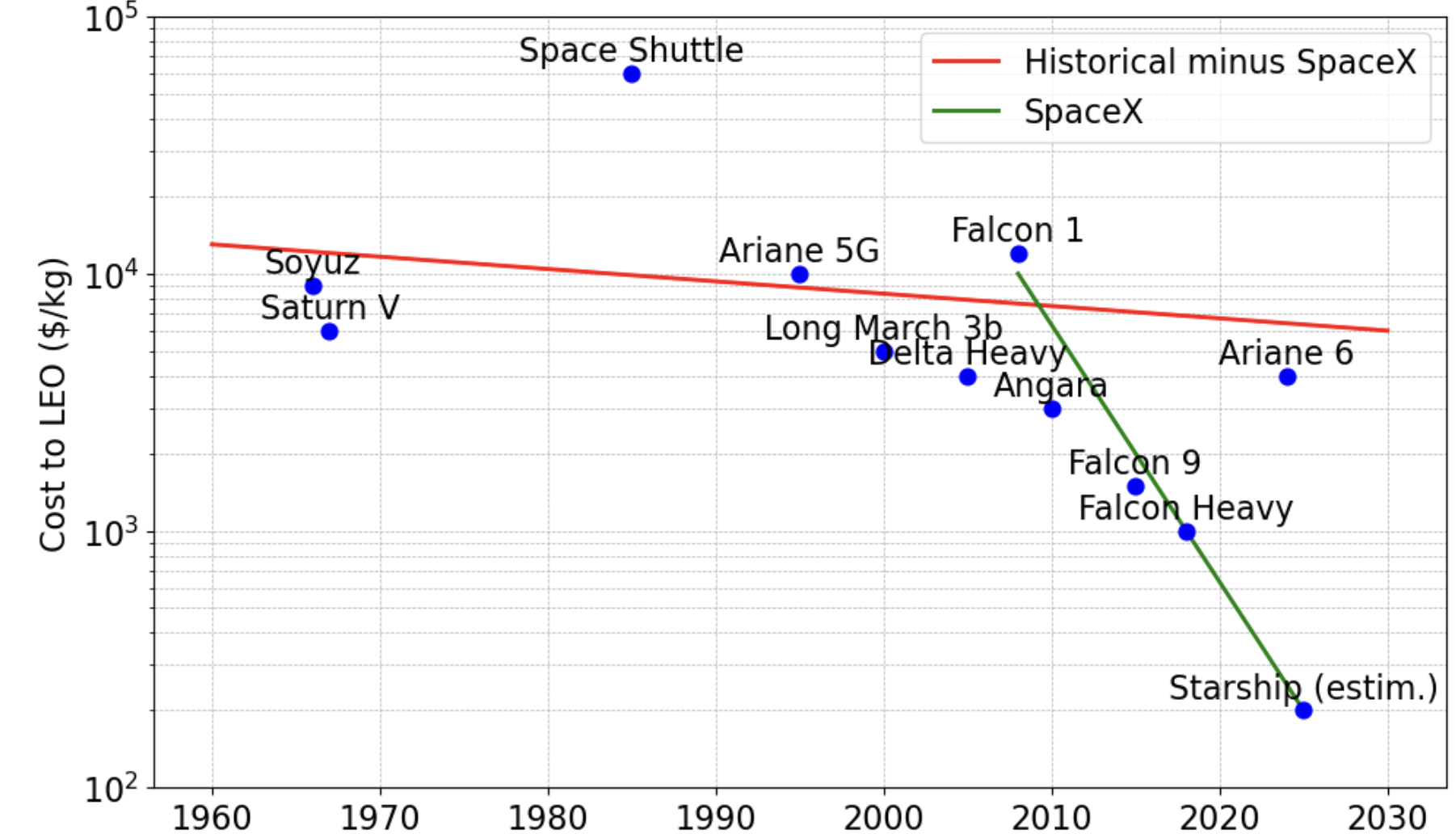}
    \caption{Cost of launch to low earth orbit. Adapted from \cite{2026-marspedia_cost_to_leo}.}
\label{fig:cost-to-leo}
\end{figure}

Figure \ref{fig:cost-to-leo} shows how the cost of launch to LEO [\$/kg] evolved over time.
The historical trendline (in red) was disrupted by SpaceX' Falcon rocket, 
resulting in the green trendline.
Reuse of the first-stage of the rocket combined with economies of scale 
reduced cost by an order of magnitude with respect to the red trendline.
With Starship (Falcon's successor) a launch-to-LEO cost of 200\$ per kg is expected.

The launch of 1GW of AI compute into LEO involves 10,000 ton of satellites\footnote{
	This is equivalent to about 50 Starship V3 launches.}
(per Musk's claim above) at a cost of 2B\$, based on the 200\$ per kg. 
These costs are modest compared to the Capex of a terrestrial AI data center, Table \ref{tab:data-centers}.


\subsection{Power supply in LEO}
\label{subsec:power}
	
Most satellites, including Starlink's, are powered by solar panels.
These solar panels can be oriented such that they are always normal to the sun.
Since LEO is well above the atmosphere the solar irradiance is high, approximately 1300  W/m$^2$.
Assuming a conversion efficiency of 30\%, these panels deliver about 400W/m$^2$.

\begin{table}[h!]
	\center
	\begin{tabular}{lrrr} 														\hline
		\textbf{Starlink} & \textbf{Mass [kg]} 	& \textbf{Area [m$^2$]} 	& \textbf{Power [kW]}  \\ \hline
		\textbf{V1} 	& $300$ 			& $25$ 				& $10$ 			\\
		\textbf{V2 mini} & $800$ 			& $100$ 				& $40$			\\
		\textbf{V3} 	& $2000$ 			& $250$ 				& $100$ 			\\ \hline 
	\end{tabular}
	\vspace{1mm}
\caption{Solar power per Starlink generation, based on 400W/m$^2$ yield.}
\label{tab:solar-power-starlink}
\end{table}	

The power available per satellite follows from the panel areas, Table \ref{tab:solar-power-starlink}
\cite{2026-wikipedia-starlink}.
Note that the 100kW of solar power (V3) is close to the power consumption
of a modern AI-compute rack, such as a 120kW NVIDIA Blackwell rack \cite{2024-nvidia-blackwell}.
	
Most LEO orbits have a day-night cycle. Presumably Starlink uses batteries for operation during nights.
For orbital AI data centers so-called sun-synchronous orbit are considered 
\cite{2025-aguera-spaceai, 2026-spacex-secfiling}.
These polar orbits provide solar energy at high irradiance, with sun-facing panels 24/7.
It is therefore fair to claim a lower OpEx for orbital AI (Table \ref{tab:data-centers}).



\subsection{Cooling}
\label{subsec:cooling}

AI compute and inter-satellite communication converts the 100kW of solar power into heat.
In space, cooling can only be by means of radiative cooling, governed by the {\em Stefan–Boltzmann law}: 
$$ P = A \!\cdot\! \sigma  \!\cdot\! T^4, ~~~ 
		\textrm{with} ~\sigma \approx 5.67 \!\cdot\!  10^{-8} ~W \!\cdot\! \!m^{-2} K^{-4} ~,
$$ 
where $A$ is the radiator area (two sides combined) and $T$ is the radiator temperature.
For simplicity, heat absorbtion by the radiator, heat transport, and emissivity are ignored.

Passive cooling assumes a highly conductive path from the heat sources (packaged compute chips)
to the radiator surfaces.
In a message on X (2026-Jan-20) Musk claimed that  ``GPUs are designed to operate at $T=370$K''. 
The Starlink V2 mini $P\!=\!40$kW at $T\!=\!370$K translates to a radiator area  $A \!\approx\! 38 m^{2}$,
or 19$m^2$ double sided.

A visual inspection of V2 mini images suggests 2 -- 3x smaller radiators.
A partial explanation could be that the solar efficiency is less than the assumed 30\%,
or that a part of the solar power is stored in batteries during daytime.
Finally, the radiative area may be increased by the use ``patterned cellular grid fin structure'',
as described in  \cite{2023-US11834205B1}.

Extending passive cooling to 100kW looks plausible.
This avoids the high CapEx of active cooling (Table \ref{tab:data-centers}),
and saves on the substantial power required for active cooling.

\subsection{Radiation Tolerance}
\label{subsec:radiation}

Satellites in LEO experience high levels of ionizing radiation, mainly energetic protons and alpha particles.
The impact of this radiation on electronic components, including compute chips is twofold.
First, there are  Total Ionizing Dose effects (TID), causing device degradation over time.
Second, there are Single Event Effects (SEE) causing bit errors.

One approach to combat these effects is radiation hardening of these components.
Starlink uses radiation hardened versions of the STM32V8 MCU (STM, 18nm SOI) 
and the Versal (AMD/Xilinx, 7nm CMOS).
Radiation hardening is costly, its availability is multiple CMOS generations behind,  and reduces performance. 
For cost effective orbital AI compute, radiation hardening is not an option.

Google proposes an alternative approach in \cite{2025-aguera-spaceai}.
They applied a 67 MeV proton beam to a packaged TPU, 
emulating radiation exposure in sun-synchronous LEO.
Using significant shielding (10 mm Al equivalent) they observed a TID dose of about 150 rad/year.
For a satellite lifetime of 5 years this corresponds to 750 rad, which the authors deemed acceptable.
Similar tests for high-bandwidth (HBM) memories were judged equally encouraging.

Single-event effects can be managed by error correcting codes,
or by error-detection and recovery procedures.


\subsection{(Unintended) Electromagnetic Radiation}
\label{subsec:electromagnetic-radiation}


From the perspective of terrestrial astronomy,
an orbital AI data center is a constellation of radiating celestial bodies.
Dominant is the infra-red emission of 100kW, which is a major headache for infrared astronomy.

\begin{figure}[t]
   \centering
    \includegraphics[width=0.9\linewidth]{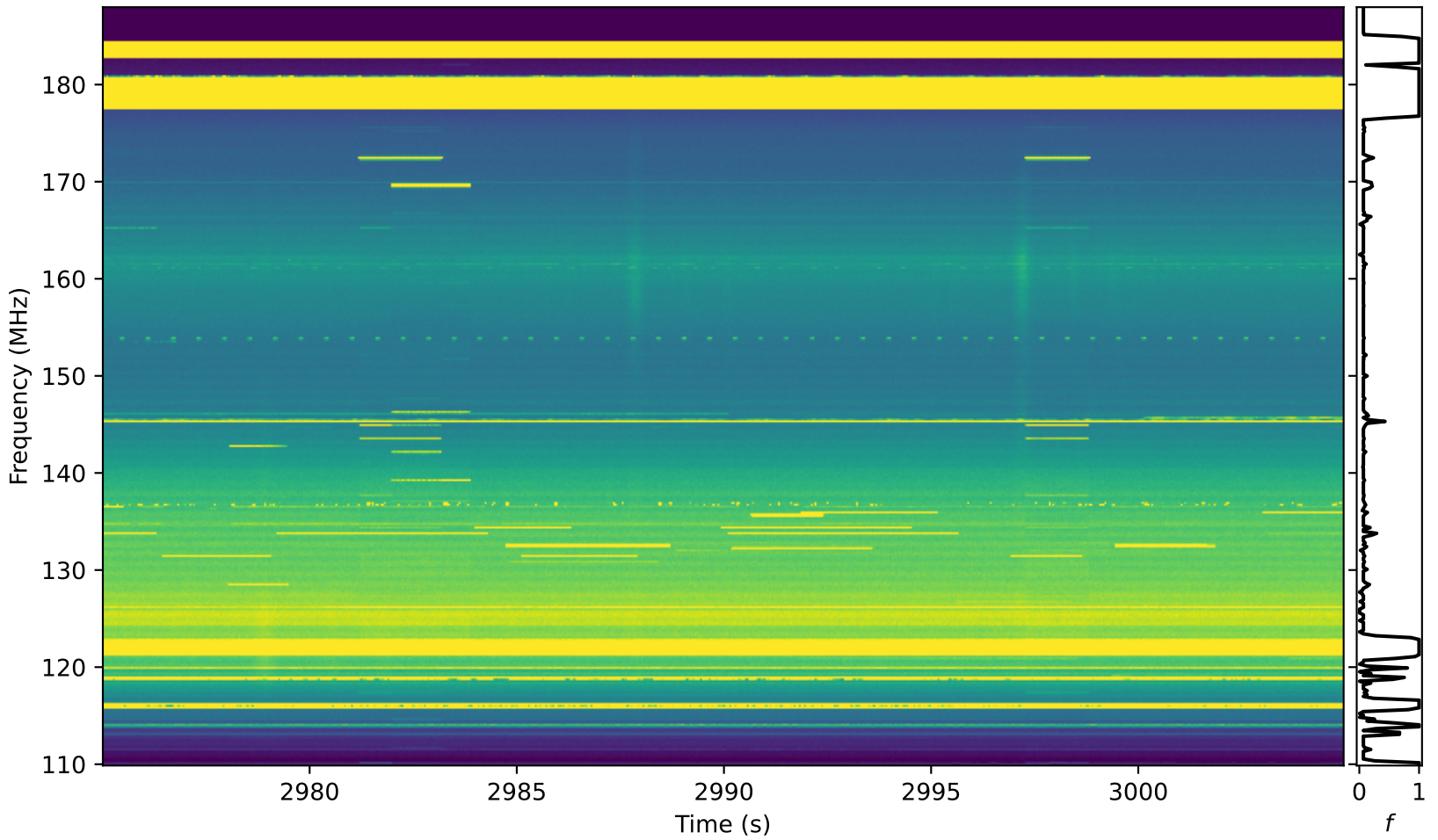}
    \caption{Unintended emission as observed by LOFAR,  from \cite{2023-lofar}.}
\label{fig:emission-lofar}
\end{figure}

There is also unintended electromagnetic radiation across the radio spectrum.
Figure \ref{fig:emission-lofar} shows the unintended radio radiation of three Starlink satellite
across a spectrum of 115 - 175MHz, during an interval of 30 seconds,
as observed by the LOFAR radio telescope \cite{2023-lofar}.
With ``Tens of thousands of LEO satellites in the making'',
``this artificial sphere of ‘radio light’ that leaks into astronomical observations, 
may render some astronomical observations impossible''.


\subsection{Reentry}
\label{subsec:reentry}

\begin{figure}[t]
   \centering
    \includegraphics[width=0.9\linewidth]{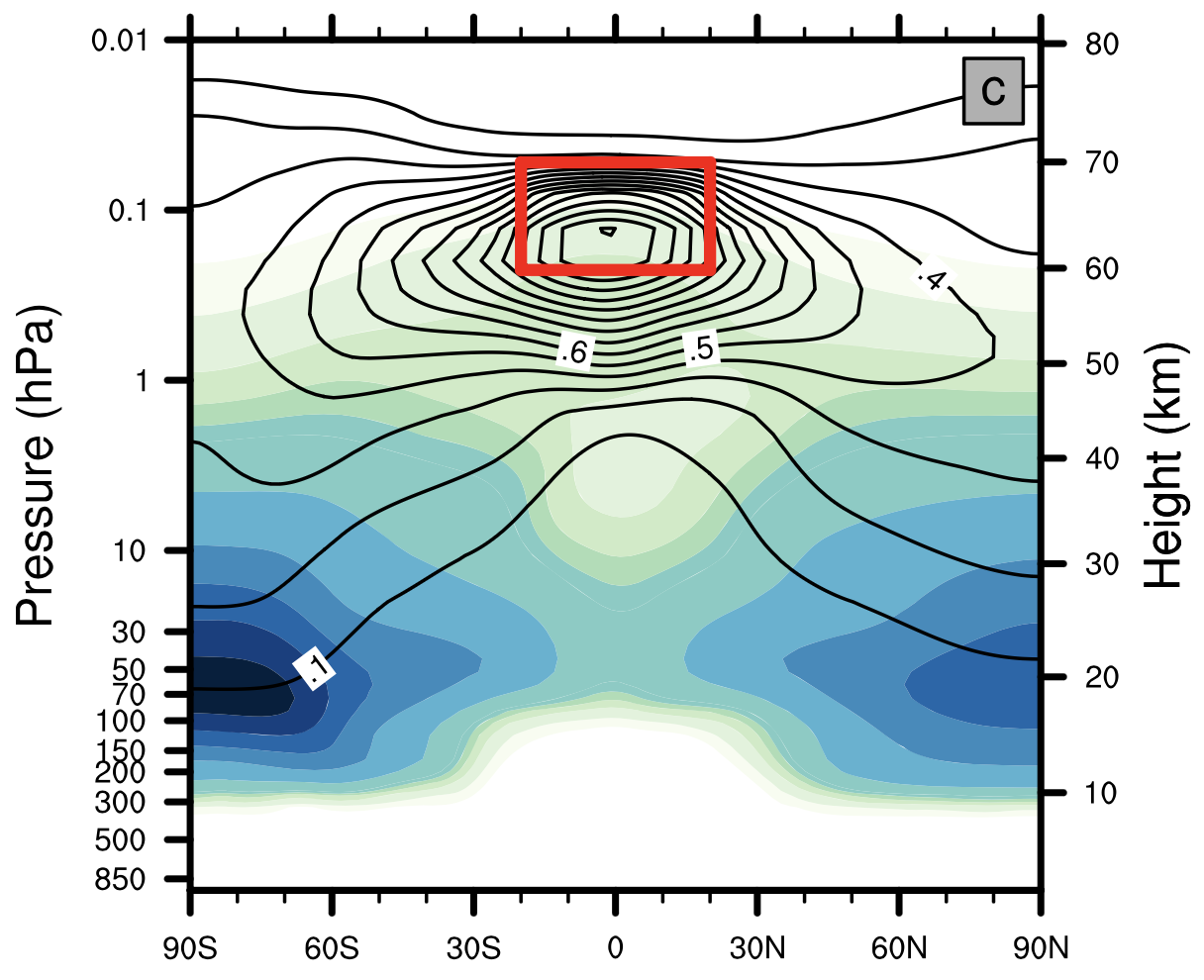}
    \caption{$Al_2O_3$ aerosol accumulation assuming satellite reentry above the equator (red box),  from \cite{2025-reentry}. }
\label{fig:reentry}
\end{figure}

LEO satellites are designed for a lifetime of about 5 years.
End-of-life implies reentry into earth' atmosphere.
The best possible outcome of such reentry is total vaporization.

The effects of such reentries has been simulated by the US NOAA \cite{2025-reentry},
assuming a total reentry mass of $10^4$ ton/year.
A satellite consists mostly of aluminium, and its vaporization results in $Al_2O_3$ aerosols,
clouds of nm-$\mu$m particulates.
Figure \ref{fig:reentry} shows that when reentry is assumed to occur at the equator,
these aerosols accumulate poleward of 30$^o$ N/S, between 10 - 30 km altitude.

These aerosols cause ``mesospheric heating rates leading to 1.5K-temperature anomalies''.
In turn, ``these anomalies lead to a weaker springtime ozone hole.''
The assumed $10^4$ ton/year is tiny compared to Musk's goal of  $10^6 - 10^7$ ton/year.


\section{The Compute Side of Orbital AI}
\label{sec:compute}

In this section we compare the training times and model-flop utilizations of frontier LLMs
running on a 1GW orbital data center versus those of the same LLM running on a 1GW terrestrial data center.
This analysis is simplified by the observation 
that a state-of-the-art compute rack consumes about 120kW \cite{2024-nvidia-blackwell}
and a state-of-the-art LEO satellite (Starlink V3) about 100kW.
From a compute perspective we consider a terrestrial rack and an orbital satellite equivalent.

In this section we develop a {\em crude} model to compare the performance
of inter-rack and inter-satellite compute networks connecting 8000 racks/satellites.


\subsection{Orbital AI Constellations and Networks }
\label{subsec:orbital-AI-networks}

Google describes a  constellation of 81 satellites flying in formation \cite{2025-aguera-spaceai}
for orbital AI applications. 
The proposed constellation is planar and has a radius of about 1km,
justified by the observation that link bandwidth scales inversely with the square of the inter-satellite distance.
The constellation shape and orientation can be maintained with only modest ``delta-v requirements'', 
a measure of the amount of propellant needed to achieve such.
For the compute network they propose free-space optical inter-satellite links,
with an  aggregate bandwidth per link, up to 10Tb/s.
They do not offer specifics on the network topology, presumably a 2D mesh.

For our study we assume two 1GW constellations of 8000 satellites each,
together with matching network topologies.
\begin{enumerate}
\item A planar constellation of $89 \!\times\! 90$ satellites, with a 2D torus as network topology.
	Each satellite has a Top of Rack (ToR) switch that connects 
	the satellite-internal network to its 4 spatial neighbors.
\item A cubic constellation of $20 \!\times\!  20 \!\times\!  20$ satellites,  with a 3D torus as network topology.
	The satellite's ToR switch connects to 6 spatial neighbors.
\end{enumerate}
Following \cite{2025-aguera-spaceai}, a inter-satellite spacing of a few 100m is assumed.
The 3D torus, with its higher-radix switches clearly offers a higher network bandwidth than the 2D torus.

There are numerous challenges with both constellations.
\begin{itemize}
\item Both constellations require lots of precise maneuvering to get the positions of the 8000 satellites roughly right.
\item The orbital dynamics of such large constellations may be more involved vs \cite{2025-aguera-spaceai}, 
 	and it may require more propellant to maintain the geometry of the constellation.
\item For the 3D constellation it is important to avoid situations where satellites cast their solar shadow on their neighbors.
\item Each satellite has 4 to 6 laser inter-satellite links.
	These $32,000$ to $48,000$ beams must be adaptively steered to the corresponding receivers. 
	This may be particularly challenging for the torus' wrap-around links.
\item One or more satellites may become unrecoverably out-of-order. 
	This requires load rebalancing and rerouting over a network that is no longer isotropic.
\item A satellite may be damaged by energetic space junk. 
	The resulting debris may result in cascades of failures among neighbor satellites, known as the Kessler syndrome.
\end{itemize}


\subsection{Bisection Bandwidths of AI Networks}
\label{subsec:bisection-bandwidths}
\newcommand{\WL}{\!\cdot\! W \!\cdot\! B_l}
\newcommand{\D}{\!\cdot\! D}

The proposed torus networks for satellite constellations introduced in \ref{subsec:orbital-AI-networks}
have vastly different network bandwidths and network latencies when compared to terrestrial networks.


{\em Bisection bandwidth} is a measure for the data volume 
a network carry per second when half the nodes send to the other half.
It is an incomplete predictor of performance for network‑bound computations,
because most communication patterns do not saturate this bisection.
For example, neighbor-to-neighbor communication patterns in a torus network
are not constrained by this bisection bandwidth.
Application of this measure is only meaningful for data volumes that do cross the bisection,
as is elaborated in \ref{subsec:bisection-intensities}.

The bisection bandwidth of a Clos network \cite{2015-google-clos} is $\frac12  N \WL $, where 
$N$ is the number of racks, 
$W$ the width of the ToR switch [\#links], and 
$B_l$ the bandwidth per link [GB/s].
This bisection width offers $\frac12  N\!\cdot\! W$ parallel non-blocking communication paths
between the $\frac12  N$ senders and $\frac12  N$ receivers.
This non-blocking property comes at price: a very large number of cables,
which can be reduced by means of oversubscription.

The bisection bandwidths of 2D and 3D toruses are $N^{1/2} \!\cdot\! B_l$ and $N^{2/3 } \!\cdot\! B_l$.
There are multiple paths between any pair of nodes, 
and a path in use between a pair of nodes, may block other communication paths. 
For training and inference algorithms the traffic patterns can be chosen such
that blocking is a non-issue.

The latency of a Clos Network is $\log(N) \cdot D$, 
where $\log(N)$ is a measure for the lengths of the  longest path between any pair of nodes. 
In practice this length is 4 to 6. $D$ is the latency per link [$\mu$s].
The latencies of 2D and 3D toruses are $N^{1/2} \D$ and $\frac{3}{2} N^{1/3} \D$  respectively.

\begin{table}[h!]
\center
\newcommand{\q}{$\!\!\!$}
\newcommand{\qq}{\q\q}
\newcommand{\qqq}{\qq\q}
\begin{tabular}{llrrr} \hline
	\textbf{network topology}\qq &   		&  \textbf{terrestrial}	& \textbf{orbital}	& \textbf{orbital}	\\ 
						&  \textbf{unit}	&  \textbf{Clos} 		& \textbf{2D torus} 	& \textbf{3D torus} 	\\ \hline
	node					& 			&  rack			& satellite			& satellite			\\
	power				& kW		& 120			& 100			& 100			\\ \hline \\[-2mm]
	$B$, bisection bw \qq 	&  &$ \qq \frac12^{\!*} N \WL$ 		& $2N^{1/2} \!\cdot\! B_l$	& $2N^{2/3 } \!\cdot\! B_l$ \\
	$W$, width ToR sw. 		& link		& $72$ 			& N.A.			& N.A.			\\      
 	$B_l$, link bw \qq		& GB/s		&  $100$	 		&  $12.5^{**}$	 	& $12.5^{**}$	 	\\
	$B$, bisection bw \qq 	& TB/s		&  28800			&  2.25			&  10  			\\
	$T_{\textit{bandwidth}}$, relative	& 	&   1 				&  12880			&  2880			\\ \hline \\[-2mm]
	$L$, network latency \qq 	&          		& $\qq\qq\log(N) \D$ &  $N^{1/2} \D$ 	&  $\frac{3}{2} N^{1/3} \D$ \\
	$D$, link latency    		& $\mu$s 		&  1   			&  20   			&  20   			\\
	$L$, network latency \qq 	& $\mu$s 		&   4				&  1800  			& 600  			\\  
	$T_{\textit{latency}}$, relative	&   		&  1				&  450			& 150			\\ \hline
\end{tabular}	
\vspace{1mm}
\caption{An overview of network bisection bandwidths and latencies for $N\!=\!8000$ racks/satellites.
		$^*$Factor $\frac12$ from a 2:1 Clos oversubscription ratio.  ~~~ $^{**}$Starlink V2 laser ISL.}
\label{tab:ai-networks}
\end{table}
These network bandwidths and latencies are summarized in Table \ref{tab:ai-networks}.
The torus topologies provide about three orders of magnitude less bisection bandwidth than a Clos network,
and the torus latencies are about two orders of magnitude larger.
$T_{\textit{bandwidth}}$ and $T_{\textit{latency}}$ refer to contributions to the overall AI-compute times
in the IMEC model described in \ref{performance-models}.


\subsection{Bisection Intensities of AI Models}
\label{subsec:bisection-intensities}

How much data actually crosses a bisection during training?
Subsection \ref{subsec:models} introduces various forms of parallelism.
During the training of a 1T-parameter model on an 8000-rack machine,
it is the {\em data parallelism} that determines the global, bisection-crossing traffic.
Other forms of parallelism (tensor, pipeline, ...) apply intra-rack, or inside a local cluster of racks.

During {\em data-parallel} LLM training, 
each rack holds a full copy of the model and processes a different slice of each global batch. 
This makes gradient sharing the only synchronization step that ties the forward and backward propagation phases together.
During a forward pass, racks run completely independently: 
each model replica computes activations for its own microbatch, and no communication is required. 
During a backward pass, each replica computes {\em local gradients} for the same parameters. 
To keep all replicas consistent, these gradients must be averaged across all racks. 

This averaging is by means of an {\em all‑reduce} operation \cite{2004-rabenseifner-allreduce, 2015-sack-torus-all-reduce}, 
in three steps:
1) every rack contributes its local gradients,  
2) the racks collectively sum them, and 
3) each rack receives the averaged result.
The inter-rack traffic involved in this all‑reduce operation crosses the network bisection.
{\em Bisection intensity} is defined as:
\begin{equation}
	I_\textit{bisect} :=  \dfrac{\textit{total FLOPs}}{\textit{data volume that crosses the bisection}} ~.
	\label{eq:bisect-def}
\end{equation}
It is a variation of {\em arithmetic intensity}, and likewise measured in [FLOP/Byte].
For LLM training it is proportional to the number of tokens per step (global batch size),
 $B_\textit{tokens}$:
\begin{equation}
	I_\textit{bisect} =  \dfrac{k}{c} B_\textit{tokens}~,
	\label{eq:bisect-expr}
\end{equation}
where $k$ denotes the FLOPs/parameter/token (typ.\ 6FLOP) and
$c$ the bytes/parameter (weights $\!+\!$ gradients $\!+\!$ related numbers, typ.\ 12Byte).
The model depth (layer count) and layer width do not show up in (\ref{eq:bisect-expr}),
as they contribute with equal factors to the numerator and denominator

The batch size $B_\textit{tokens}$ increases with model size, and is bounded by training convergence considerations.
For models around 1T-parameters, $B_\textit{tokens} \approx 1$M is typical. 
This corresponds with an $I_\textit{bisect} \approx 500$k: 
for every byte of data crossing the bisection, all racks together deliver $500$k of useful FLOPs.


\subsection{Rooflines for AI Inference}
\label{subsec:ai-inference}

Rooflines were introduced as a tool to analyze the performance 
of systems that are constrained by compute [\#flops] and by communication [\#bytes/sec].
Whereas rooflines are commonly used to analyze memory traffic along a memory hierarchy,
they can applied to network traffic as well.

In contrast to AI training, AI inference involves only a few racks, for seconds to minutes.
For a 1T-parameter model, 
deploying 8 model replicas (for load balancing),
using 64$\times$ batching (for weight reuse),
and aiming at low latency (for interactive use), a handful of racks would do.

\begin{figure}[t]
   \centering
    \includegraphics[width=0.9\linewidth]{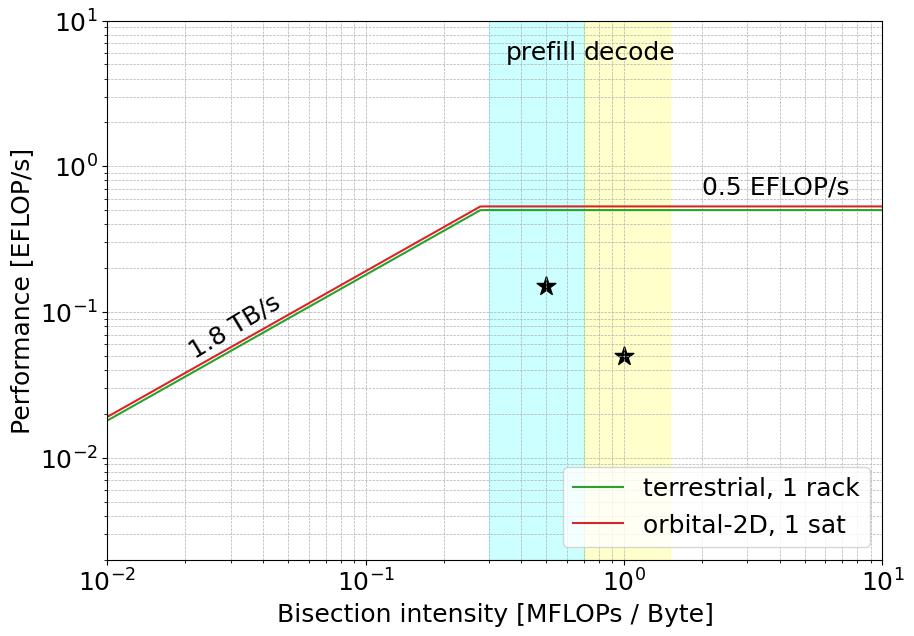}
    \caption{The rooflines for terrestrial and orbital inference of a 1T-parameter LLM
    		are the same when inference is  confined to a single rack/satellite.
		The bisection bandwidth applies to {\em intra-rack} communication.}
\label{fig:roofline-inference}
\end{figure}

It is a reasonable goal for orbital inference to confine the workload to a single satellite.
In the near term this can likely be achieved by increasing the compute efficiency [FLOP/Joule].
Figure \ref{fig:roofline-inference} offers the rooflines for single-satellite/rack inference.

The inference workload is split into two steps: prefill and decode.
For both steps an indicative range of bisection intensities is given (colored boxes).
The actual prefill and decode performances numbers (the stars) are well below these rooflines,
because both steps are not network bound but memory bound  \cite{2022-mfu-inference-google, 2026-imec-performance-model}.


\subsection{Rooflines for AI Training}
\label{subsec:ai-training}

\begin{figure}[t]
   \centering
    \includegraphics[width=0.9\linewidth]{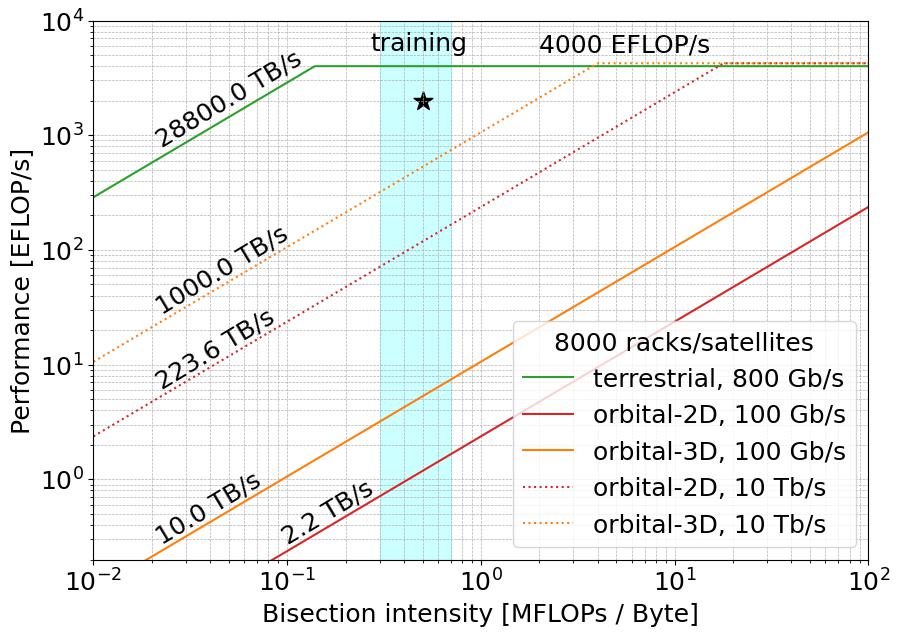}
    \caption{Rooflines for terrestrial/orbital training of a 1T-parameter LLM,
    		assuming 8000 racks/satellites, for specified link bit rates.
		(The dotted lines represent 10Tb/s FSO links, discussed in Section \ref{sec:discussion}.)}
\label{fig:roofline-training}
\end{figure}

Figure \ref{fig:roofline-training} shows rooflines for the training a 1T-parameter model,
running on a terrestrial 8000-rack data center, and on the two 8000-satellite
constellations introduced in \ref{subsec:orbital-AI-networks}.
The bisection bandwidths are from Table \ref{tab:ai-networks},
and the blue-colored range of bisection intensity values is from 
\ref{subsec:bisection-intensities}.

For terrestrial training the network bandwidth is not the limiting factor,
and therefore the assumed Clos oversubscription ratio of 2:1 appears rather conservative.
The star corresponds with the estimated bisection intensity of 500kFLOPs/Byte
and a performance of 2000 EFLOP/s,
matching  the earlier reported  model flop utilizations of  $\approx50\%$.

For orbital training the low bisection bandwidths of the torus networks reduce 
the performance by two to three orders of magnitude.
It is even worse: the most common algorithms used for all-reduce on torus networks 
\cite{2015-sack-torus-all-reduce}
require the all-reduce data volume to cross the bisection multiple times, once per dimension.
These reductions in performance correspond to matching reductions in model flop utilizations
and translate to corresponding increases in orbital training costs.


\section{Discussion}
\label{sec:discussion}

In Oct. 2025 Musk stated \cite{2025-xai-joins-spacex}:
``My estimate is that within 2 to 3 years, the lowest-cost way to generate AI compute will be in space.'' 
His perspective: lower construction costs, lower costs of (solar) power, and lower costs of (passive) cooling.
Presumably he also factors in the high costs of terrestrial regulation, NIMBY, long lead times, property taxes, etc.

The costs imposed on the astronomy community are not part of his calculation (\ref{subsec:electromagnetic-radiation}).
Neither are the costs related to the impact on the earth ozone layer 
resulting from the reentry of millions of tons of aluminium into earth's atmosphere, each year (\ref{subsec:reentry}).

The prospects for cost-effective LLM inference in space look promissing, 
provided inference is performed on individual satellites optimized for that task (\ref{subsec:ai-inference}).
Each individual satellite handles a stream of prompts. 
There is no need for the complex, dense satellite constellations described in \ref{subsec:orbital-AI-networks}.

The prospects for cost-effective LLM training in space look daunting (\ref{subsec:ai-training}).
The costs of network communication among thousands of satellites is overwhelming,
resulting in very low model flop utilizations and excessive training times.
These findings are based on a rather crude model, 
viz.\ a roofline model using network bisection bandwidths and bisection intensities.
These bandwidths in turn assume speculative planar and cubic satellite constellations
with matching torus networks, and 100Gb/s laser inter-satellite links.
Accordingly, a $100^+$ times increase in training costs is merely a rough indication.

What are the options to address these excessive training costs? Below we discuss four:
1) larger satellites with denser compute, 2) innovations in free-space optics,
3) ``Clos satellite constellations'', and 4) radically new ways of parallel training.

An increase in the compute capacity per satellite by, say, $8\times$ would reduce
an $N= 20 \!\times\! 20 \!\times\! 20$ satellite constellation into a $10 \!\times\! 10 \!\times\! 10$ one.
However, from a network communication point-of-view this is counter productive,
as the number of links crossing a bisection plane reduces from $20 \!\times\! 20$ to $10 \!\times\! 10$, 
due to scaling with $N^{2/3}$.

Developments in free-space optics offer a potential game changer \cite{2026-fso-overview,2025-aguera-spaceai}.
Aggressive application of Wavelength Division Multiplexing (WDM) and spatially-multiplexed beams may
increase link bandwidths dramatically,
from a Starlink-V2 100Gb/s to an envisioned 10Tb/s \cite{2025-aguera-spaceai}.
The impact on the orbital rooflines is shown by the dotted lines in Fig. \ref{fig:roofline-training}.
Beam stearing then becomes more challenging, especially for the torus wrap-around links.
Even if all this could be realized, the resulting network would still be inferior to a terrestrial Clos network,
in part due to higher latencies incurred by the many hops in a torus (Table \ref{tab:ai-networks}).

A ``Clos satellite constellation'' would mimick the topology of a Clos network.
It would consists of compute satellites plus a number of communication satellites
that implement the leaf, spine,  and super‑spine switches \cite{2015-google-clos}, appropriately clustered.
There are various geometries possible for such a constellation.
A "bi-planar" version would consists of one plane with 8000 compute satellites,
and a parallel plane with all the communication satellites.
All links presumably use free-space optics, with an appropriate numbers of beams per link.

Radical innovations on parallel training could help if they increase the bisection intensity by some factor
(Fig.\ \ref{fig:roofline-training}). Unfortunately, there are no obvious alternatives to all-reduce.

The SpaceX SEC filing \cite{2026-spacex-secfiling} is somewhat cautious:
``The timeline for deploying 100 gigawatts of annual compute power to orbit
involves unproven or new innovations, and may be difficult or impossible to determine''.
This provision does not question the feasibility itself, but merely the difficulty of determining the timeline.

Cost-effective training of frontier LLMs in space would require overcoming many challenges,
including constellation formation and control, 
coordinated steering of many thousands of WDM laser beams, 
and robust adaptation to satellite and link failures. 
Furthermore, even if FSO eventually delivers on its potential on link bandwidths, 
network-wide latencies inherent to large orbital meshes
may make the earlier cited terrestrial model flop utilizations of 40-70\% unrealistic.

Given the scale and difficulty of these obstacles,
cost-effective frontier-LLMs training in space within the next 2-3 years is not credible
and even a 10-year horizon appears optimistic with current technology trends.

\subsection*{Acknowledgements}

The author thanks Rudolf Mak and Kees van Hee (both TU/e Eindhoven), 
Wayne Burleson and Mortaza Hassini (both Univ. of Massachusetts Amherst), and John Voris
for their careful reviews of an earlier draft.


\bibliographystyle{IEEEtran}
\bibliography{AI-in-The-Sky.bib}

\end{document}